# High-Precision Megahertz-to-Terahertz Dielectric Spectroscopy of Protein Collective Motions and Hydration Dynamics

Ali Charkhesht,[1] Chola K. Regmi,[1] Katie R. Mitchell-Koch,[2] Shengfeng Cheng,[1,3] Nguyen Q. Vinh[1,*]

[1] Department of Physics and Center of Soft Matter and Biological Physics, Virginia Tech, Blacksburg, Virginia 24061
[2] Department of Chemistry, Wichita State University, Wichita, Kansas 67260
[3] Macromolecules Innovation Institute, Virginia Tech, Blacksburg, Virginia 24061

Corresponding author: *Email: vinh@vt.edu; Phone: 540-231-3158

**ABSTRACT:** The low-frequency collective vibrational modes in proteins as well as the protein-water interface have been suggested as dominant factors controlling the efficiency of biochemical reactions and biological energy transport. It is thus crucial to uncover the mystery of hydration structure and dynamics as well as their coupling to collective motions of proteins in aqueous solutions. Here we report dielectric properties of aqueous bovine serum albumin protein solutions as a model system using an extremely sensitive dielectric spectrometer with frequencies spanning from megahertz to terahertz. The dielectric relaxation spectra reveal several polarization mechanisms at the molecular level with different time constants and dielectric strengths, reflecting the complexity of protein-water interactions. Combining the effective-medium approximation and molecular dynamics simulations, we have determined collective vibrational modes at terahertz frequencies and the number of water molecules in the tightly-bound and loosely-bound hydration layers. High-precision measurements of the number of hydration water molecules indicate that the dynamical influence of proteins extends beyond the first solvation layer, to around 7 Å distance from the protein surface, with the largest slowdown arising from water molecules directly hydrogen-bonded to the protein. Our results reveal critical information of protein dynamics and protein-water interfaces, which determine biochemical functions and reactivity of proteins.

## I. INTRODUCTION

Biological functions of proteins in aqueous environments, such as enzymatic activity, oxygen transport, neuron signal transmission, and ion channels for signaling currents, depend on their structural changes, flexibility, and protein-water interface[1-7]. Specifically, protein flexibility and vibrational modes have been considered to be responsible for efficiently directing biochemical reactions and biological energy transport. It has been suggested that low-frequency collective vibrational modes (<3 THz) involving dynamical networks extending throughout the protein play a crucial role in controlling the structural changes. As a general rule, biological functions of proteins occur in aqueous environments[8]. While solvation effects on proteins play an essential role in the structure, stability, and dynamics of proteins, our understanding of solvent dynamics at the protein-water interface remains inadequate. Water molecules that populate the surfaces of proteins[9-11], lipid bilayers[12], lipid headgroups[13], or in the crowded milieu of tissues and cells[14] exhibit properties that are particularly important to the structure and biological functions of a protein, and distinct from those found in the bulk. With a small size and a large dipole moment, water molecules form stable layers around proteins to perform a multitude of functions in biological environments. The structure and dynamics of these layers, which are determined by hydrophobic and hydrophilic interactions, among other influences, are important for biological functions. Our understanding of the flexibility of a protein and how its environment contributes to catalytic mechanisms lags far behind our knowledge of three-dimensional structures and chemical mechanisms.

A wide range of experimental and computational techniques have been employed to investigate the molecular dynamics of proteins in solution.[15] Different techniques may provide information on different aspects of protein dynamics. For example, some techniques, such as nuclear magnetic resonance (NMR)[6, 16] and Mössbauer spectroscopy[17-18], are used to detect the motion of probe nuclei, while others including optical pumping[19-20], femto-second pump-probe[21-23], optical Kerr-effect[11, 24], and inelastic neutron scattering experiments[25-26] report on the dynamics of more globally distributed probes. Some experiments (e.g., inelastic neutron scattering, NMR) require complex facility-based methods as wells as cryogenic temperatures and/or non-physiological conditions. The measured timescales of protein motions also differ between techniques[1]. X-ray crystallography collects average atomic positions over hours; Mössbauer spectroscopy probes motions on a timescale of about $10^{-7}$ s; neutron scattering detects motions on timescales ranging from $10^{-12}$ to $10^{-8}$ s, depending on instrumental resolution. Molecular dynamics (MD) simulations[27-30] are largely limited by accessible timescales which at best can reach $10^{-6}$ s (though some recent MD simulations of protein dynamics can reach $10^{-3}$ s timescale but require special hardware). One approach, megahertz (MHz) to terahertz (THz) dielectric spectroscopy, provides observations that are relevant to the global and subglobal motions on the picosecond to sub-microsecond timescales.[10, 13, 31-32] The dielectric spectroscopy is a label-free technique that is carried out at physiological conditions of proteins. However, far-infrared and terahertz time-domain spectroscopy (THz-TDS) have been limited by the very large absorption of liquid water. Although recent THz-TDS experiments on lysozyme crystals have successfully identified underdamped delocalized vibrational motions in the terahertz frequencies, the protein dynamics are strongly affected by crystal packing[2].

The properties of hydration water such as dielectric susceptibility and relaxation frequency are strongly affected by the slow but large-scale motion of a protein molecule[10, 13, 33]. However, using THz radiation, several groups reported different results for the effects of proteins on water structure. For example, Ding et al.[34] estimated a hydration layer with a thickness of ~11–17 Å from the peptide surface using THz-TDS experiments. They argued that effects of a peptide surface in an aqueous solution are beyond the first hydration shell, but no long-range effect on water structure has been identified. Employing a p-germanium laser at 2.25 and 2.55 THz, Ebbinghaus et al.[35-36] reported a nonlinear behavior of THz absorption with protein concentration, which was attributed to the onset of overlapping dynamical hydration layers, leading to conclusions that a protein has a long-range effect on water beyond 20 Å from the protein surface. The suggestion became controversial as it was in contrast with previous findings of MD simulations[27], protein functional studies[37], magnetic relaxation dispersion measurements[38], and high precision densitometry experiments[39]. These aforementioned works concluded that proteins dominate only one hydration layer rather than multiple hydration layers. Recently, on the basis of "hydration layer overlap" hypothesis, Bye et al.[40] used a modified calculation model to explain the variation in absorption coefficients with protein concentration. However, this model does not take the absorption of proteins in the THz frequencies into consideration, which makes the explanation lose its universality. The controversy may also originate from distinct definitions of hydration layers, and the sensitivity of the apparatus. Given the controversy resulting from these works, it is important to investigate the validity of the explanation of the measurements.

Advances in MHz-to-THz technology make it possible to conduct a more thorough study of the dielectric response of proteins in an aqueous environment. Such a study can act as an important step in understanding motions of complex biomolecular systems at timescales from microsecond to picosecond. Here, we report the dielectric response from MHz-to-THz spectroscopy of a benchmark protein, bovine serum albumin (BSA), in aqueous solutions. Through our developments we have achieved a very large spectral bandwidth from MHz to THz frequencies and improved the signal-to-noise ratio by several orders of magnitude, providing continuous high-fidelity coverage that no other techniques can match[41]. Employing high-precision measurement, we systematically study the nature of biological water and the associated dynamics of proteins at the molecular scale.

## II. EXPERIMENTAL METHODS

Our spectrometers cover a broadband spectral range from MHz to THz frequencies that allows us to observe both the relaxational (rotational) and translational processes of water molecules (bulk, tightly- and loosely-bound water) as well as biomolecules. Figure 1a shows the absorption and refractive index of BSA solutions for several solutions (see materials and methods in Supplemental Material). With simultaneous measurements of the absorption and refractive index, the complex refractive index, $n^*(\nu)$, of a material can be expressed as a function of frequency, $\nu$:

$$n^*(\nu) = n(\nu) + i\kappa(\nu) \qquad (1)$$

where $n(\nu)$ is the refractive index of the solution, and $\kappa(\nu)$ is the extinction coefficient of the solution and is related to the absorption coefficient, $\alpha(\nu)$, by $\kappa(\nu) = c\alpha(\nu)/(4\pi\nu)$ with $c$ being the speed of light. Similarly, the complex dielectric response of a material can be expressed as:

$$\epsilon^*_{\text{sol}}(\nu) = \epsilon'_{\text{sol}}(\nu) + i\eta''_{\text{sol}}(\nu) \qquad (2)$$

where $\epsilon'_{\text{sol}}(\nu)$ and $\eta''_{\text{sol}}(\nu)$ are the real (dielectric dispersion) and imaginary (total dielectric loss) components of the relative permittivity, respectively. The imaginary component of the dielectric response of BSA solutions contains two contributions:

$$\eta''_{\text{sol}}(\nu) = \epsilon''_{\text{sol}}(\nu) + \eta''_\sigma(\nu)$$
$$= \epsilon''_{\text{sol}}(\nu) + \sigma/(2\pi\nu\epsilon_0) \qquad (3)$$

where $\epsilon''_{\text{sol}}(\nu)$ represents the dielectric loss, $\eta''_\sigma(\nu)$ is the Ohmic loss due to the drift of ions contained in BSA solutions, $\sigma$ is the electrical conductivity of the solution, and $\epsilon_0$ is the permittivity of the vacuum. The electrical conductivity of BSA solutions has been obtained at the frequency of 1 kHz and at 25 °C (Figure S1, Supplemental Material). These values have been used for the extraction of the dielectric loss of BSA solutions from the total dielectric loss. Since we simultaneously measure both the absorption and refractive index of materials, the complex dielectric response of BSA solutions can be calculated from the following relations:

$$\epsilon'_{sol}(\nu) = n^2(\nu) - \kappa^2(\nu) = n^2(\nu) - (c\alpha(\nu)/4\pi\nu)^2$$
$$\epsilon''_{sol}(\nu) = 2n(\nu) \cdot \kappa(\nu) = 2n(\nu)c\alpha(\nu)/4\pi\nu$$
(4)

The complex dielectric response spectra (dielectric dispersion, $\epsilon'(\nu)$, and the dielectric loss, $\epsilon''(\nu)$, components) of water and BSA solutions at 25 °C have been extracted from the absorption and refractive index measurements (Figure 1b). A main dielectric loss peak frequency centered at ~19 GHz remains virtually unchanged. This dielectric loss peak has been attributed to bulk water in the solution[13, 42-43]. However, the presence of BSA proteins in solutions induces a pronounced broadening of dielectric loss spectra on the lower frequency side of the main dielectric loss peak. The dielectric response for the orientational relaxation of molecular dipoles with the size of 66 kDa (BSA protein) is in the range of 1 to 10 MHz[13, 43]. Thus, the low-frequency broadening is not likely due to relaxation processes of bulk water as well as BSA proteins in the solution, but can be attributed to the emergence of new slow relaxation modes.

## III. RESULTS AND DISCUSSION

**Megahertz to Gigahertz Spectroscopy.** Dielectric relaxation spectrometry of aqueous BSA solutions at MHz to GHz frequencies provides insight into mechanisms of reorientational dynamics of water. The technique is definitive for inferring the distinctly different dynamics of water molecules in hydration shells and those in the bulk (Figure 1).

At the microscopic level, several dielectric mechanisms or polarization effects contribute to the dielectric response of aqueous solutions. Water molecules as well as biomolecules with permanent dipoles rotate to follow an alternating electric field. In this frequency range, three contributions dominate in the dielectric response of aqueous protein solutions, including (*i*) the rotational motion of biomolecules, i.e., orientational relaxation of biomolecular dipoles; (*ii*) the orientational polarization of bulk water molecules, i.e., water dipoles in the bulk; (*iii*) the orientational polarization of water molecules in the interfacial region surrounding biomolecules, i.e., water dipoles in hydration shells. Atomic and electronic polarizations are relatively weak, and usually constant over the MHz to GHz frequency.

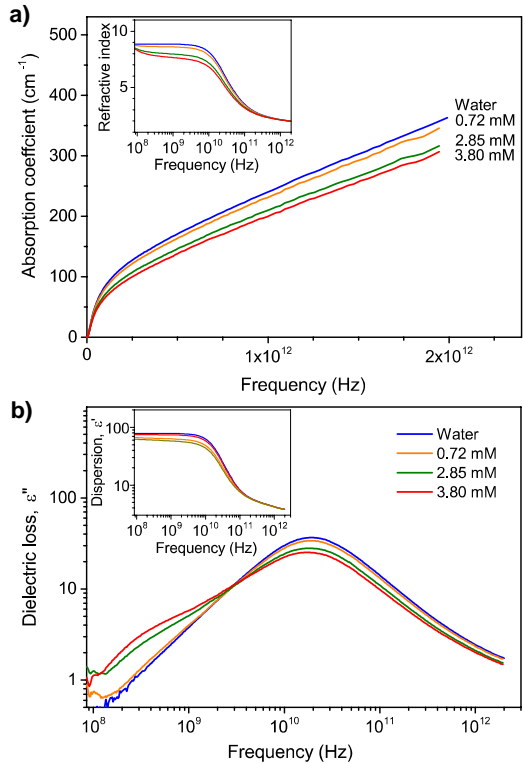

**Figure 1:** The interaction of MHz to THz radiation and BSA proteins providing the dynamics over picosecond to sub-microsecond timescales. (**a**) The MHz to GHz absorption of both BSA solutions and water increases monotonically with rising frequency at 25 °C. The refractive indices (upper inset) of BSA solutions and water decrease with increasing frequency. (**b**) The dielectric loss, $\epsilon''(\nu)$ and the dielectric dispersion spectra, $\epsilon'(\nu)$, in the lower inset, from BSA solutions and water were obtained from the absorption and refractive index measurements. The main dielectric loss peak frequency centered at ~19 GHz remains unchanged. An addition of BSA proteins in solutions produces a pronounced broadening on the lower frequency side of the dielectric loss spectra.

Dielectric spectra of BSA solutions in the low frequency region (below 50 GHz) are shown in Figure 2. Typically, the dielectric response of the orientational relaxation of permanent dipoles from macromolecules with a molecular weight of ~66 kDa is in the range of 1 to 10 MHz[13, 43]. We do not focus on this frequency range in this paper. The dielectric response of the orientational polarization of bulk water centered at ~19 GHz has been well established[42, 44-45]. However, the contribution of the dielectric response of water in hydration layers of proteins is complex and less understood. For example, the dielectric response of hydration water of ribonuclease A[46], lysozyme protein[43], and zwitterionic dodecylphosphocholine micelles[13] in aqueous solutions consists of several dispersion regions. Typically, in a simple approximation, hydration water can be classified into two types including tightly-bound water (with dielectric response in 100 – 500 MHz) and loosely-bound water (with dielectric response in 1 – 5 GHz). The tightly-bound water molecules have direct and strong contacts with biomolecular surfaces. We have shown that water molecules having direct but weak interaction with biomolecular surfaces (such as with a soft cation) can be categorized as loosely-bound water[13]. Thus, loosely-bound hydration water include water molecules in outer hydration shells and water molecules having weak interactions with biomolecular surfaces, regardless of location. These loosely-bound water molecules exchange with tightly-bound water and have dynamics approaching those of bulk water.

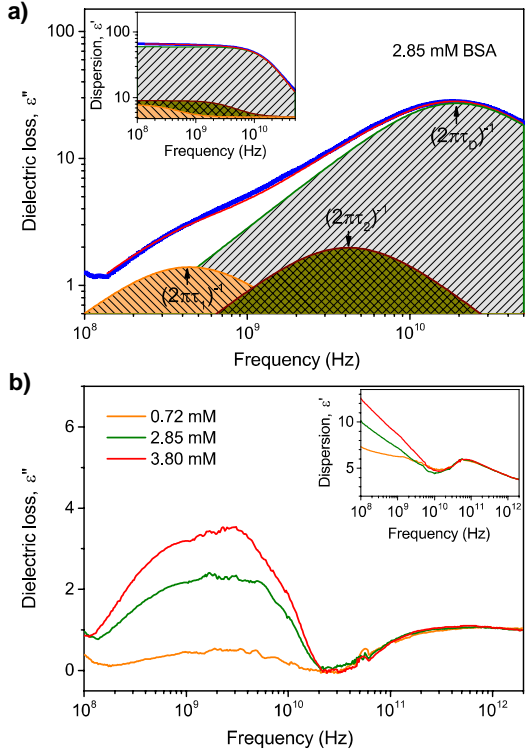

**Figure 2:** The dielectric response of BSA aqueous solutions in the frequency range from 100 MHz to 50 GHz showing the heterogeneity on a scale of several water layers around proteins. (**a**) Dielectric spectra for both dielectric dispersion (upper inset), $\epsilon'_{sol}(\nu)$, and dielectric loss, $\epsilon''_{sol}(\nu)$, together with their spectral deconvolution provide insight into the dynamics of water molecules at the protein surface for the 2.85 mM BSA solution. The red curves are fits of the real and imaginary components of the complex dielectric response. (**b**) The dielectric loss and dielectric dispersion (lower inset) spectra of tightly- and loosely-bound water for several BSA solutions have been obtained by subtracting the well-defined relaxation contribution of bulk water from the total spectra. The procedure reveals the distinctly different dynamic behavior of hydration layers compared to bulk water.

In the MHz to GHz frequency region, the data were analyzed by simultaneously fitting dielectric dispersion and dielectric loss components to relaxation models based on the sum of *n* individual contributions described by the Havriliak-Negami function[47]. The librational motions and inertial effects do not contribute appreciably to the dielectric response, yielding a usually constant ($\epsilon_\infty$) value over the region. Thus, the dielectric response is in the form:

$$\epsilon^*(\nu) = \epsilon_\infty + \sum_{j=1}^{n} \frac{\epsilon_{j-1}-\epsilon_j}{\left(1+(i2\pi\nu\tau_j)^{1-\alpha_j}\right)^{\beta_j}} \quad (5)$$

Each mode is characterized by an amplitude (relaxation strength), $\Delta\epsilon_j = \epsilon_j - \epsilon_{j+1}$, a relaxation time, $\tau_j$ ($\tau_j > \tau_{j+1}$), and shape parameters, $0 \leq \alpha_j < 1$ and $0 < \beta_j \leq 1$. The simplified variants include the Cole-Davidson ($\alpha_j = 0; 0 < \beta_j \leq 1$), the Cole-Cole ($0 < \alpha_j \leq 1; \beta_j = 1$), and the Debye ($\alpha_j = 0; \beta_j = 1$) equations. In equation (5), $\epsilon_0 = \epsilon_s$ is the static permittivity which can be defined as $\epsilon_s = \epsilon_\infty + \sum_{j=1}^{n} \Delta\epsilon_j$, and $\epsilon_n = \epsilon_\infty$ is the dielectric constant at infinity frequency that captures contributions of dielectric response from modes with frequencies much higher than the probed range, and thus reflects contributions from atomic and electronic polarizations. We have fitted equation (5) simultaneously to the measured dielectric dispersion, $\epsilon'_{sol}(\nu)$, and the dielectric loss, $\epsilon''_{sol}(\nu)$, to minimize errors. Our analyses show that it is sufficient to consider Debye-type relaxations. Specifically, our data are reasonably fitted with a superposition of three Debye relaxation processes in the form:

$$\epsilon^*_{sol}(\nu) = \epsilon_\infty + \frac{\epsilon_S-\epsilon_1}{1+i2\pi\nu\tau_1} + \frac{\epsilon_1-\epsilon_2}{1+i2\pi\nu\tau_2} + \frac{\epsilon_2-\epsilon_D}{1+i2\pi\nu\tau_D} \quad (6)$$

where $\Delta\epsilon_1 = \epsilon_S - \epsilon_1$, $\Delta\epsilon_2 = \epsilon_1 - \epsilon_2$ and $\Delta\epsilon_D = \epsilon_2 - \epsilon_\infty$ are dielectric strengths of each Debye process to the total relaxation from tightly-bound, loosely-bound, and bulk water, respectively, with $\tau_1$, $\tau_2$, and $\tau_D$ as the corresponding relaxation times. It is important to note that, at 25 °C, the dielectric spectra of pure water could be formally fitted by a sum of three Debye processes[42, 44], yielding $\tau_D \sim 8.27$ ps (19 GHz), $\tau_{D2} \sim 1.1$ ps (145 GHz), and $\tau_{D3} \sim 0.178$ ps (895 GHz). In the frequency range up to 50 GHz, the contributions of the dielectric response of bulk water from $\tau_{D2}$ and $\tau_{D3}$ modes are small and included in the $\varepsilon_\infty$ parameter.

Using this approach, we fit both the dielectric dispersion, $\epsilon'_{sol}(\nu)$, and loss, $\epsilon''_{sol}(\nu)$. Six parameters in equation (6) were varied simultaneously, while the relaxation time for bulk water, $\tau_D \sim 8.27$ ps, was held fixed at the literature value[42, 44, 48-49]. Dielectric spectra for both dielectric dispersion, $\epsilon'_{sol}(\nu)$, and dielectric loss, $\epsilon''_{sol}(\nu)$, together with their spectral deconvolution are shown in Figure 2a for a 2.85 mM

BSA solution. The dielectric loss spectrum indicates three relaxation processes centered at 441 ± 23 MHz ($\tau_1 \sim 361 \pm 19$ ps), 4.19 ± 0.85 GHz ($\tau_2 \sim 38 \pm 11$ ps), and 19.25 ± 0.78 GHz ($\tau_D \sim 8.27 \pm 0.35$ ps) for tightly-bound, loosely-bound, and bulk water, respectively. Corresponding to the three relaxation processes in the dielectric loss spectrum, the dielectric dispersion is shown in the inset of Figure 2a. Our fitting to the three-Debye relaxational model in the low frequency region produces $\varepsilon_\infty = 5.2 \pm 0.5$ for the 2.85 mM BSA solution. This value is within experimental uncertainty of the prior literature values[44, 49]. We have also obtained the values of the dielectric strength, with $\Delta\varepsilon_1 = 2.8 \pm 0.3$ for tightly-bound, $\Delta\varepsilon_2 = 3.9 \pm 0.3$ for loosely-bound, and $\Delta\varepsilon_D = 54.3 \pm 0.3$ for bulk water, respectively. The dominant contribution from the $\tau_D$ relaxation process reflects the cooperative reorientational dynamics of dipole moments of bulk water in the protein solution.

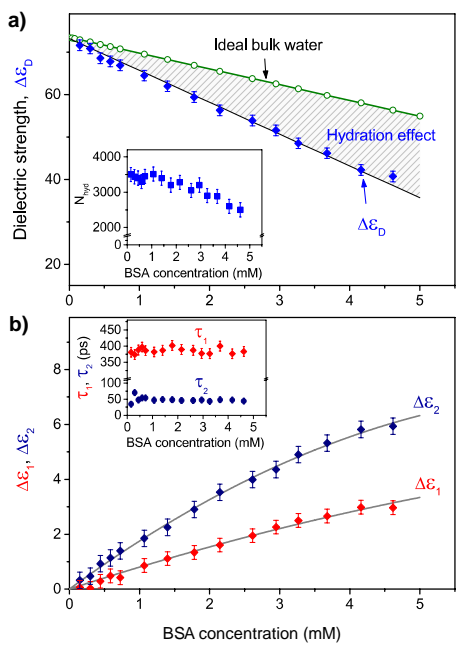

**Figure 3:** Dielectric relaxation measurements showing the existence of several relaxation processes in protein solutions. (**a**) The dielectric strength of the bulk water, $\Delta\varepsilon_D$, in BSA solutions significantly decreases with increasing protein concentration. The continuous solid line (green) represents the dielectric amplitude of ideal bulk water calculated under an assumption that all water molecules in solution behave as bulk water and participate in the relaxation process at ~19 GHz. The hydration number, $N_{hyd}$, as a function of protein concentration (upper inset) deduced from the dielectric strength provides the number of water molecules that do not participate in the relaxation process of bulk water because of the hydration effect. (**b**) Amplitudes of the dielectric response of the tightly- and loosely-bound water in solutions increase with increasing protein concentration. The relaxation time constants (lower inset) of water in hydration shells are constant with protein concentrations.

The dielectric measurements provide insights into the dynamics of water at the protein interface and the heterogeneous nature of hydration shells at a molecular level. Amplitudes of relaxation processes and relaxation time constants deduced from fitting experimental spectra to the three-Debye model (equation 6) are shown in Figure 3. Specifically, the relaxation times are independent of the protein concentration, $c$, over the entire concentration range that we have explored (Figure 3b, inset). Two long relaxation time constants, 379 ± 22 ps and 45 ± 11 ps, for the reorientation of water molecules in hydration shells have been identified, in addition to the relaxation time of bulk water of 8.27 ps. The dynamics of the two slower processes are slower than that of bulk water relaxation by factors of 47 and 6, respectively. The amplitudes of all three relaxation processes, in contrast, vary strongly with protein concentration. While the dielectric strength of bulk water, $\Delta\epsilon_D$, smoothly decreases with increasing protein concentration (Figure 3a), those of the two slower processes, $\Delta\epsilon_1$ and $\Delta\epsilon_2$, increase when $c$ increases. Giving further insight into the amplitudes of the two slower relaxation processes, we have observed that their dielectric strengths show saturation at high protein concentrations.

The two slower processes originate from the cooperative molecular dynamics of water molecules in hydration layers. In these layers, water molecules are densely packed, and their orientations are thus highly cooperative, leading to a detectable slowdown in their dielectric relaxation times compared to that of bulk water. The longest relaxation time, $\tau_1 = 379 \pm 22$ ps, comes from water molecules in the tightly-bound hydration layer. These water molecules have strong and direct hydration bonds with the protein surface. The loosely bound water molecules, having indirect contacts or weaker hydration interactions with the protein surface, relax with the intermediate relaxation time, $\tau_2$, of 45 ± 11 ps.

The protein hydration structure is not only quite thin, compared to the size of the protein, but also heterogeneous on the scale of several water layers around the protein. To focus on the dielectric response of water in hydration layers, the dielectric spectra of the bound water in the MHz to GHz frequencies have been obtained by subtracting the contribution of bulk water in solution characterized by relaxation time, $\tau_D$, and the dielectric strength, $\Delta\varepsilon_D$, from the total dielectric response, $\epsilon^*_{sol}(\nu)$. Using this approach, we have obtained dielectric response of tightly-bound and loosely-bound hydration water for several BSA solutions (Figure 2b). This procedure reveals that dispersion and loss curves clearly exhibit relaxation processes of hydration water. The dielectric response at THz frequencies of BSA solutions is complex, including the dielectric response of collective motions of BSA proteins, and the librational and vibrational processes of water. As discussed below, we have analyzed the dielectric response of protein solutions at high frequencies separately, using the effective-medium approximation and MD simulations.

Information on the hydration volume, including the number of water molecules affected by the protein, can be obtained from the weighted contribution of each relaxation process. The addition of proteins to water alters the structure and dynamics of surrounding water molecules. This process leads to a cooperative rearrangement of the hydration-bond network of bulk water. The lowering of the amplitude of the dielectric response for bulk water with increasing protein concentration comes from two main factors: (*i*) the presence of proteins and (*ii*) water molecules in hydration shells. With the presence of proteins, the concentration of water in the solution is lower than the pure water, thus, reducing the dielectric response. We have calculated the dielectric response of the solution under the assumption that all water molecules in the solution participate in the $\tau_D$ relaxation process as in pure water, which is the "ideal bulk water" (the continuous green line in Figure 3a). However, the dielectric response of bulk water, $\Delta\epsilon_D$, in the protein solution from our measurements is lower, indicating that not all water molecules in the solution relax via the $\tau_D$ relaxation mode. As discussed above, the water molecules in hydration shells have much longer relaxation times. When the protein concentration increases, the fraction of hydration water increases, resulting in a lower dielectric response of the solution. For a given protein concentration, we employ this method to estimate the number of water molecules residing in all hydration shells by comparing the measured dielectric response of bulk water in the protein solution to the calculated dielectric value when all water molecules in the solution are treated as bulk water. This method has been used frequently in literature[13, 43, 46, 50], and the number of water molecules in hydration shells per protein, termed hydration number, is given by:

$$N_{\text{hyd}}(c) = \frac{c_\text{w} - \frac{\Delta\epsilon_\text{w}}{\Delta\epsilon_\text{pure}} c_\text{pure}}{c} \quad (7)$$

where $c_\text{w}$ is the molar concentration of water in the solution, and $c_\text{pure}$ = 55.35 M is the molarity and $\Delta\epsilon_\text{pure}$ = 73.25 is the dielectric strength of pure water at 25 °C.[42, 51]

We have calculated the number of water molecules per protein that no longer participate in the relaxation process of bulk water due to the hydration effect. Our analyses show that $N_{\text{hyd}}$ = 3500 ± 200 for BSA solutions when the protein concentration is lower than 3 mM, and then starts to decrease as the protein concentration increases further (Figure 3a, inset). When the protein concentration is low, the solution is dilute and the average distance between proteins is much larger than the thickness of hydration shells. In this case the hydration shell is solely determined by water-protein interactions, and the hydration number, $N_{\text{hyd}}$, is thus, to the first approximation, independent of protein concentration. The estimation of $N_{\text{hyd}}$, indicates that less than 3 layers of water molecules surrounding the protein surface are affected by the protein at low concentrations. Using MD simulations below, we have estimated about 4100 water molecules within 7.0 Å from the protein surface, which is less than 3 layers of water molecules around protein. When the protein concentration increases to a certain level, hydration shells start to overlap, and proteins aggregate in small equilibrium clusters, resulting in a decrease of the hydration number. A similar trend of the hydration number has been reported in dielectric measurements of lysozyme[43] and micelles[13] in a wide range of concentrations. The small-angle neutron scattering for DPC micelles has also indicated a decrease of micellar hydration number with increasing concentration as surfactants become more closely packed and compete with one another for space within the micellar arrangement[52].

**Terahertz Spectroscopy.** With a higher frequency, terahertz radiation has been used to probe vibrational modes typically involving collective atomic motions of macromolecules, which include both inter- and intramolecular interactions. Terahertz spectroscopy of biomolecules in aqueous environments, thus, provides an important approach for identifying their global and transient molecular structures as well as directly assessing hydrogen-bonding and other detailed environmental interactions[10]. However, a significant challenge in obtaining terahertz dielectric spectra of aqueous biomolecular solutions is the strong absorption of water in the spectral range of 0.5 – 10 THz. Using our high resolution and high dynamic terahertz frequency-domain spectroscopy, we have determined the absorption and refractive index spectra of protein solutions along with that of pure water (Figure 1). The absorption and the refractive index data indicate strong frequency dependence, increasing and decreasing with increasing frequency, respectively. It has been shown that the most prominent effect of adding solutes to water is a monotonic decrease in absorption with increasing solute concentration[2, 10, 53]. This is primarily due to the fact that the highly absorbing solvent is replaced by the solutes having lower absorption at THz frequencies.

As frequently discussed in the literature, the influence of protein-water interactions extends beyond the tightly-bound hydration layer, causing changes to the hydrogen-bonding network, and thus resulting in a strong dependence of the THz absorption on protein concentration[10, 35, 40, 43, 54-55]. As a first approximation, we can consider protein solutions as a two-component system with BSA proteins and surrounding bulk water molecules. The total absorption of the solution as a weighted average of its two constituents is given by:

$$\alpha_{sol} = (\alpha_{wat}V_{wat} + \alpha_{BSA}V_{BSA})/V_{sol} \qquad (8)$$

where $\alpha_{sol}$, $\alpha_{wat}$ and $\alpha_{BSA}$ are absorption coefficients of the solution, water, and BSA protein, while $V_{sol}$, $V_{wat}$ and $V_{BSA}$ are their volumes, respectively.

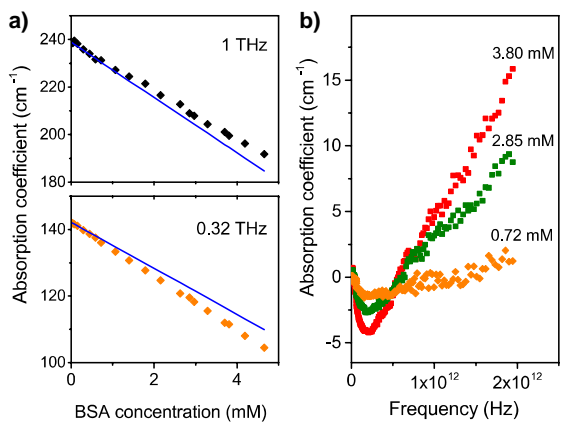

**Figure 4:** The THz absorption of hydrated BSA provides the low-frequency vibrational dynamics of proteins in water. (**a**) The THz absorption coefficient of BSA without any correction for the protein's hydration shell (absorption of water in solutions is subtracted from solution absorption) reveals negative absorption, which is unphysical. Data points represent the experimental data, whereas solid lines show the calculated absorption reduction due to the exclude volume of the protein. Specifically, the absorption of BSA proteins in solutions shows negative absorption at 0.32 THz (lower) and positive absorption at 1 THz (upper). (**b**) The absorption spectra of BSA proteins in water show negative absorption in the range from 50 GHz to 650 GHz for several protein concentrations.

Following this approach, the absorption of proteins in a solution can be determined by subtracting the absorption of water from the total spectrum. We have measured the absorption coefficient of BSA solutions with concentrations from 0.038 mM (dilute) to almost the saturating level of 4.656 mM, at different frequencies of 0.32 THz and 1 THz (Figure 4a). During the sample preparation, we determined accurately the volume of water in protein solutions, thus we can calculate the absorption coefficient of water in protein solutions (solid lines in Figure 4a). As can be seen from Figure 4a, the measured absorption is lower at 0.32 THz and higher at 1 THz than the calculated absorption of water in BSA solutions. It means that the absorption of BSA proteins in water is negative at 0.32 THz and positive at 1 THz. In order to confirm the measurements, we have measured the absorption of BSA solutions as a function of frequency. The absorption of BSA proteins in water turns out to be negative in a frequency window from 50 GHz to 650 GHz for several protein concentrations (Figure 4b). A similar observation has been reported for lysozyme in water[10].

This unphysical result has previously been explained by the fact that the hydrophilic surface of a protein interact with water and bind some water molecules adjacent to them[10]. Therefore, the water molecules having strong hydrogen bonds with the protein surface cannot take part in the dipole relaxation, or in other words, they do not contribute to the absorption of bulk water in the THz frequency window. We note that the binding to the hydrophilic surface of water molecules does not necessarily mean that these water molecules are immobile. Instead, these water molecules form a tightly-bound hydration shell around a protein, and their dynamics are slower than those of bulk water molecules. Assuming the minimum of the curves in Figure 4b is zero, we can estimate the number of water molecules removed from bulk water. Using this method, the "*lost*" solvent corresponds to 750 ± 75 water molecules bound to a BSA protein (equivalent to 0.20 ± 0.02 g of water per gram of BSA protein). This value is less than a single layer of water molecules that fully cover the surface of a BSA protein. The changes of absorption at the THz frequencies with BSA concentrations follow the "Beer's law", in which the measured absorbance is proportional to the protein concentration. The observation suggests that the number of water molecules in the tightly-bound hydration layer of a BSA protein is independent of protein concentrations over the entire range investigated here. The assumption that the minimum absorbance of proteins is precisely zero may be not convincing, thus the above estimation yields the lower-bound of water molecules in the tightly-bound hydration shell.

In dealing with a highly heterogeneous system, we have reported both absorption and refractive index or the complex dielectric response of protein solutions with high precision, rather than assuming that overall absorption of a solution is the sum of the absorption of its constituents. The analysis based on only absorbance measurements, as we have noted above, may fail in spectral regions for which the refractive index of the solvent changes very rapidly with frequency (Figure 1a). With simultaneous measurements of the absorbance and refractive index, we can apply effective-medium methods to extract the dielectric response of the solute: hydrated BSA proteins. A protein solution is a mixture of water and protein, each of which possesses its own complex dielectric constant, $\varepsilon^*_{\text{wat}}(\nu)$ and $\varepsilon^*_{\text{pro}}(\nu)$, respectively. The complex dielectric response of the protein solution, $\varepsilon^*_{\text{sol}}(\nu)$ has been determined from the experimental observables of absorption and refractive index, $\alpha$ and $n$, respectively (Figure 1b). We assume that (*i*) proteins in solution are spherical with a radius of $R_p$ and have a spherical shell containing tightly-bound water with a thickness of $d$; (*ii*) water molecules in the tightly-bound hydration shell are a part of the hydrated protein; and (*iii*) water outside the hydrated proteins has the same dielectric property as that of pure, bulk water. Note that the spherical shell of tightly-bound water can be a fraction of one monolayer of water molecules around the protein. Because the size of hydrated proteins is orders of magnitudes smaller than the wavelengths of the incoming electromagnetic radiation, the material can be treated as a homogeneous substance with an effective-dielectric function using the effective-medium approximation of Bruggemann[56] which effectively treats both low and high concentration mixtures. (Note: the Maxwell Garnet[57], and the Wagner-Hanai approximations[58-59] are for the low concentration limit of the Bruggeman approximation). The composite medium is determined from:

$$f_{\text{hp}} \frac{\varepsilon^*_{\text{hp}} - \varepsilon^*_{\text{sol}}}{\varepsilon^*_{\text{hp}} + 2\varepsilon^*_{\text{sol}}} + \left(1 - f_{\text{hp}}\right) \frac{\varepsilon^*_{\text{wat}} - \varepsilon^*_{\text{sol}}}{\varepsilon^*_{\text{wat}} + 2\varepsilon^*_{\text{sol}}} = 0 \qquad (9)$$

where $\varepsilon^*_{\text{hp}}$ is the complex dielectric response of hydrated proteins described as the process of adding water to dry proteins; $\varepsilon^*_{\text{wat}}$ is the complex dielectric response of water; $f_{\text{hp}} = (N_p/V)(4\pi/3)(R_p + d)^3$ is the volume fraction of hydrated proteins, and $N_p/V$ is the concentration of proteins in solution. The information provides insights into the protein dynamics as well as the number of water molecules in the tightly-bound hydration shell.

Employing the Bruggemann effective-medium analysis, we found that each BSA protein captures a tightly-bound hydration shell composed of 1150 ± 95 water molecules. Unlike the absorbance-based method used above and in prior literature[35, 40, 54], this method of estimating the size of the tightly-bound hydration shell requires only the well-founded assumption that the protein's absorption falls to zero at zero frequency[10]. The value of 1150 water molecules for the THz-defined hydration shell corresponds to a sub-monolayer on the surface of a BSA protein. Specifically, if we approximate BSA as a sphere with a diameter ~4 nm, then a single layer of water fully covering its surface contains ~1500 water molecules. Using the number of water molecules in the tightly-bound hydration shell related to the scaled filling factor and the measured $\varepsilon^*_{\text{wat}}(\nu)$ and $\varepsilon^*_{\text{sol}}(\nu)$, we have obtained the dielectric spectra of hydrated BSA proteins in several solutions (Figure 5). It should be noted that the hydration number calculated from the effective-medium approximation is lower than that from the GHz dielectric relaxation measurements which determine the total number of water molecules affected by the protein. This is expected as the effective-medium approximation yields the number of water molecules that have strong hydrogen bonds with the protein. These water molecules become an integral part of the protein and cannot move easily (but are not completely immobile); they form the tightly-bound hydration layer.

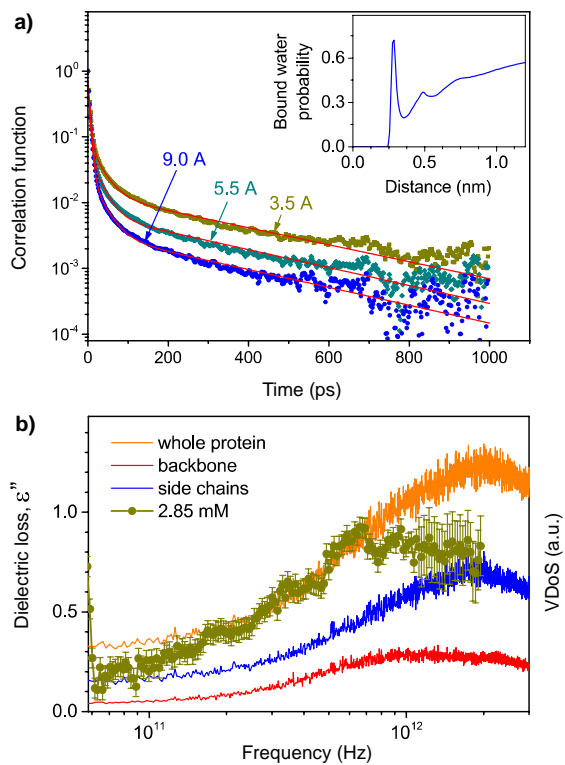

**Figure 5:** Dielectric spectra of hydrated BSA proteins in the THz frequencies and rotational autocorrelation functions, $P_1(t)$, of water providing insight into the collective motions of hydrated proteins and the dynamics of water molecules around protein surfaces. (**a**) Rotational autocorrelation functions of water molecules within 3.5, 5.5 and 9.0 Å from protein surfaces indicate three distinct dynamics corresponding to those of bulk water, tightly- and loosely-bound water around proteins, respectively. The solvent radial distribution function (upper inset) allows us to extract the number of water molecules in the tightly-bound hydration shell of a hydrated protein. (**b**) The dielectric loss spectrum (dark yellow symbols) of hydrated BSA proteins at 25 ºC is extracted from the effective-medium approximation. The VDoS calculations for the side chains (blue curve), backbone (red curve), and whole protein (orange curve) have a broad peak at 1.6 THz.

**Molecular Dynamics Simulations.** To provide a molecular-level picture of the dynamics and structure of hydration shells as well as the collective motions of proteins in solution, we have conducted MD simulations for BSA proteins in water. The combination of MD simulations with the MHz to THz spectroscopy leads to a microscopic level understanding of the coupled protein-water dynamics.

The MD simulations were performed using GROMACS package (version 5.1.4) with cubic periodic boundary conditions. The gromos54a7 force field and SPC/E water model were employed. The system was equilibrated during the first 200 ps in NPT ensemble at constant pressure (1 bar) and temperature (298 K) using Parrinello-Rahman barostat and V-rescale thermostat, and then followed by a 40 ns trajectory in a constant NVT (canonical) ensemble. The equations of motion were integrated with a time step of 2 fs, and trajectories were saved every 20 fs. Bond lengths were constrained using LINCS algorithm. Columbic and Lennard-Jones interactions were truncated at 1.4 nm. Long-range electrostatic interactions were calculated using particle mesh Ewald method with an order of 4. GROMACS tools were used for data analyses. The vibrational density of states (VDoS) in the THz frequency range was calculated from the last 2 ns of trajectories using the velocity autocorrelation tool.

The tightly-bound hydration shell of proteins can be identified by examining the average density of water molecules around the protein surface. In particular, we consider oxygen and nitrogen atoms on the protein surface and calculate the density of water molecules as a function of the water-protein distance. The result is a radial density function (RDF) of water (Figure 5a, inset). From the water RDF, we are able to extract the number of water molecules in the hydration shells of a BSA protein. The first peak of the RDF is at 3.5 Å, while the second peak occurs around 5.5 Å. By taking a time-average of the number of water molecules within the first peak of the RDF, we have determined that the number of water molecules having strong interactions with protein surface is $1230 \pm 75$. This value is in excellent agreement with the number of water molecules ($1150 \pm 95$) in the tightly-bound hydration shell estimated from the dielectric spectroscopy at THz frequencies using the effective-medium approximation. Using the same method, we can estimate $4100 \pm 75$ water molecules within 7.0 Å from the protein surface.

To resolve the dynamics and the total volume of water affected by the protein, we have employed GROMACS tools to analyze the rotational (reorientation) autocorrelation functions of water molecules. Such analyses can provide information of hydration shells including how far out into the solvent the influence of the protein can reach, and how the water dynamics is affected. Using simulation data for trajectories that are 40 nanosecond long, rotational autocorrelation functions of water using blocks of 1 ns were calculated and averaged. To monitor the dynamics of water around the protein at a large distance from the protein, we have performed analyses for water within different thicknesses from the protein surface. The rotational autocorrelation functions based on the first Legendre polynomial ($P_1$) for water molecules within 3.5, 5.5 and 9.0 Å from the protein surface, respectively, are given in Figure 5a.

The rotational autocorrelation functions of water around proteins can be fitted as superpositions of three exponential decay processes. Specifically, relaxation time constants of $6.85 \pm 0.85$, $39 \pm 9$ ps, and $335 \pm 50$ ps were obtained for all three thicknesses, respectively. The amplitudes of the three processes, in contrast, vary strongly with the water thickness. The fastest relaxation time arises due to the bulk water in the volume around protein, that has been observed previously by nuclear magnetic resonance (NMR)[60], pump-probe experiments[61], and our dielectric relaxation spectroscopy[42]. The slower relaxation times of $335 \pm 50$ ps and $39 \pm 9$ ps align well with the experimental values of 379 and 45 ps for tightly- and loosely-bound water, respectively. These water molecules relax more slowly because of their hydrogen bonds and other intermolecular interactions with the protein surface[62]. When the thickness of water from the protein surface is larger than 9 Å or 3 water layers from the protein surface, the contribution to the rotational autocorrelation function mainly comes from bulk water. This finding is in agreement with our estimation for the total number of water molecules (~$3500 \pm 200$) affected by the protein in dielectric response measurements.

The contribution of the collective dynamics of BSA proteins in an aqueous solution to the THz spectra can be obtained by computing the vibrational density of states (VDoS) spectra using MD data (Figure 5b). The VDoS calculations for the side chains (blue curve), the backbone (red curve), and the whole protein (i.e., the side chains and the backbone; orange curve) all show a broad peak at 1.6 THz, which is consistent with the experimental spectra (dark yellow symbols). The analyses allow us to delineate contributions from different motions. The MD data show that the motion of the protein side chains has a larger contribution to the density of states than the backbone, although both increase with frequency, as is observed in the experimental THz spectra. The higher flexibility of side chains in this frequency window agrees with MD calculations for lysozyme proteins in water using analysis of root mean square fluctuations (RMSFs)[11]. Thus, our analyses reveal that the collective motions in this frequency window primarily arise from side-chain torsional oscillations or hindered motions about terminal single bonds[11]. Both the THz spectrum and the backbone VDoS have maximum intensity around 1 THz, which supports the assertion that THz dielectric spectroscopy is sensitive to the large-domain motions of proteins in solution.

## IV. CONCLUSION

Employing our high-precision dielectric spectroscopy in a wide frequency range, from 100 MHz to 2 THz, and MD simulations, we have demonstrated that vibrational spectra of BSA proteins in water exhibit a broad peak at 1.2 THz, and water molecules in hydration shells of the protein reveal retarded reorientation dynamics relative to bulk water. The MHz to GHz dielectric spectroscopy reveals three main relaxational processes of water in aqueous BSA solutions, including the tightly-bound water with a reorientation time of 379 ps, loosely-bound water with a relaxation time of 45 ps, and bulk water with a relaxation time of 8 ps. The hydration number or the total amount of hydration water molecules affected by the presence of a BSA protein has been determined to be ~3500, including both tightly- and loosely-bound water. When the protein concentration is higher than 3 mM, hydration shells start to overlap, and proteins aggregate in small equilibrium clusters, resulting in a decrease of the hydration number. Using the effective-medium approximation for the dielectric measurements at the THz frequencies, we are able to extract the number of tightly-bound water molecules, which is about 1150 molecules per protein, as well as the collective vibrational modes of hydrated BSA proteins. MD simulations yield results in excellent agreement with experiments. In particular, simulations show that there are ~1230 water molecules directly hydrogen-bonded to the surface of a BSA protein, and the total hydration layer has a mean thickness of ~7 Å. Our results indicate that the effects of a BSA protein in water is beyond the first hydration shell but there is no long-range effect on bulk water structure. MD simulations also reveal that the dominant contribution to the THz peak in the dielectric spectra comes from large-domain motions of BSA proteins.

**Supplemental Material.** The materials, solution preparation as well as absorption, refractive index and electrical conductivity, σ, measurements of aqueous BSA solutions.

## ACKNOWLEDGEMENTS

N.Q.V. and K.R.M. acknowledge the support from NSF (CHE-1665157). C.R. and S.C. acknowledge the support from The Thomas F. and Kate Miller Jeffress Memorial Trust, Bank of America, N.A., Trustee via the Jeffress Trust Awards Program in Interdisciplinary Research. Acknowledgment is made to the Donors of the American Chemical Society Petroleum Research Fund for partial support of this research. This work was also supported by a grant from the Institute of Critical Technology and Applied Sciences (ICTAS) at Virginia Tech.

**REFRENCE:**

1. Daniel, R. M.; Dunn, R. V.; Finney, J. L.; Smith, J. C. The Role of Dynamics in Enzyme Activity. *Annu Rev Bioph Biom* **2003**, *32*, 69-92.
2. Acbas, G.; Niessen, K. A.; Snell, E. H.; Markelz, A. G. Optical Measurements of Long-Range Protein Vibrations. *Nat Commun* **2014**, *5*, 3076.
3. Bahar, I.; Rader, A. J. Coarse-Grained Normal Mode Analysis in Structural Biology. *Curr Opin Struc Biol* **2005**, *15*, 586-592.
4. Sundstrom, V. Light in Elementary Biological Reactions. *Prog Quant Electron* **2000**, *24*, 187-238.
5. Tan, M. L.; Bizzarri, A. R.; Xiao, Y. M.; Cannistraro, S.; Ichiye, T.; Manzoni, C.; Cerullo, G.; Adams, M. W. W.; Jenney, F. E.; Cramer, S. P. Observation of Terahertz Vibrations in Pyrococcus Furiosus Rubredoxin via Impulsive Coherent Vibrational Spectroscopy and Nuclear Resonance Vibrational Spectroscopy - Interpretation by Molecular Mechanics. *J Inorg Biochem* **2007**, *101*, 375-384.
6. Wand, A. J. Dynamic Activation of Protein Function: A View Emerging from NMR Spectroscopy. *Nat Struct Biol* **2001**, *8*, 926-931.
7. Yu, X.; Leitner, D. M. Anomalous Diffusion of Vibrational Energy in Proteins. *J Chem Phys* **2003**, *119*, 12673-12679.
8. Levy, Y.; Onuchic, J. N. Water mediation in protein folding and molecular recognition. *Annu Rev Bioph Biom* **2006**, *35*, 389-415.
9. Bellissent-Funel, M. C.; Hassanali, A.; Havenith, M.; Henchman, R.; Pohl, P.; Sterpone, F.; van der Spoel, D.; Xu, Y.; Garcia, A. E. Water Determines the Structure and Dynamics of Proteins. *Chem Rev* **2016**, *116*, 7673-7697.
10. Vinh, N. Q.; Allen, S. J.; Plaxco, K. W. Dielectric Spectroscopy of Proteins as a Quantitative Experimental Test of Computational Models of Their Low-Frequency Harmonic Motions. *J Am Chem Soc* **2011**, *133*, 8942-8947.
11. Turton, D. A.; Senn, H. M.; Harwood, T.; Lapthorn, A. J.; Ellis, E. M.; Wynne, K. Terahertz Underdamped Vibrational Motion Governs Protein-Ligand Binding in Solution. *Nat Commun* **2014**, *5*, 3999.
12. Ho, C.; Slater, S. J.; Stubbs, C. D. Hydration and Order in Lipid Bilayers. *Biochemistry-Us* **1995**, *34*, 6188-6195.
13. George, D. K.; Charkhesht, A.; Hull, O. A.; Mishra, A.; Capelluto, D. G. S.; Mitchell-Koch, K. R.; Vinh, N. Q. New Insights into the Dynamics of Zwitterionic Micelles and Their Hydration Waters by Gigahertz-to-Terahertz Dielectric Spectroscopy. *J Phys Chem B* **2016**, *120*, 10757-10767.

14. Ball, P. Water as an Active Constituent in Cell Biology. *Chem Rev* **2008,** *108*, 74-108.
15. Mondal, S.; Mukherjee, S.; Bagchi, B. Protein Hydration Dynamics: Much Ado about Nothing? *J Phys Chem Lett* **2017,** *8*, 4878-4882.
16. Ishima, R.; Torchia, D. A. Protein Dynamics from NMR. *Nat Struct Biol* **2000,** *7*, 740-743.
17. Frolov, E. N.; Gvosdev, R.; Goldanskii, V. I.; Parak, F. G. Differences in the Dynamics of Oxidized and Reduced Cytochrome C Measured by Mossbauer Spectroscopy. *J Biol Inorg Chem* **1997,** *2*, 710-713.
18. Schunemann, V.; Winkler, H. Structure and Dynamics of Biomolecules Studied by Mossbauer Spectroscopy. *Rep Prog Phys* **2000,** *63*, 263-353.
19. Suomivuori, C. M.; Gamiz-Hernandez, A. P.; Sundholm, D.; Kaila, V. R. I. Energetics and Dynamics of a Light-Driven Sodium-Pumping Rhodopsin. *P Natl Acad Sci USA* **2017,** *114*, 7043-7048.
20. Inoue, K.; Ono, H.; Abe-Yoshizumi, R.; Yoshizawa, S.; Ito, H.; Kogure, K.; Kandori, H. A Light-Driven Sodium Ion Pump in Marine Bacteria. *Nat Commun* **2013,** *4*, 1678.
21. Jumper, C. C.; Arpin, P. C.; Turner, D. B.; McClure, S. D.; Rafiq, S.; Dean, J. C.; Cina, J. A.; Kovac, P. A.; Mirkovic, T.; Scholes, G. D. Broad-Band Pump-Probe Spectroscopy Quantifies Ultrafast Solvation Dynamics of Proteins and Molecules. *J Phys Chem Lett* **2016,** *7*, 4722-4731.
22. Armstrong, M. R.; Ogilvie, J. P.; Cowan, M. L.; Nagy, A. M.; Miller, R. J. D. Observation of the Cascaded Atomic-to-Global Length Scales Driving Protein Motion. *P Natl Acad Sci USA* **2003,** *100*, 4990-4994.
23. Karunakaran, V.; Denisov, I.; Sligar, S. G.; Champion, P. M. Investigation of the Low Frequency Dynamics of Heme Proteins: Native and Mutant Cytochrome P450(cam) and Redox Partner Complexes. *J Phys Chem B* **2011,** *115*, 5665-5677.
24. Hunt, N. T.; Kattner, L.; Shanks, R. P.; Wynne, K. The Dynamics of Water-Protein Interaction Studied by Ultrafast Optical Kerr-Effect Spectroscopy. *J Am Chem Soc* **2007,** *129*, 3168-3172.
25. Lerbret, A.; Hedoux, A.; Annighofer, B.; Bellissent-Funel, M. C. Influence of Pressure on the Low-Frequency Vibrational Modes of Lysozyme and Water: A Complementary Inelastic Neutron Scattering and Molecular Dynamics Simulation Study. *Proteins* **2013,** *81*, 326-340.
26. Roh, J. H.; Curtis, J. E.; Azzam, S.; Novikov, V. N.; Peral, I.; Chowdhuri, Z.; Gregory, R. B.; Sokolov, A. P. Influence of Hydration on the Dynamics of Lysozyme. *Biophys J* **2006,** *91*, 2573-2588.
27. Bizzarri, A. R.; Cannistraro, S. Molecular Dynamics of Water at the Protein-Solvent Interface. *J Phys Chem B* **2002,** *106*, 6617-6633.
28. Comez, L.; Paolantoni, M.; Sassi, P.; Corezzi, S.; Morresi, A.; Fioretto, D. Molecular Properties of Aqueous Solutions: a Focus on the Collective Dynamics of Hydration Water. *Soft Matter* **2016,** *12*, 5501-5514.
29. Heyden, M.; Tobias, D. J. Spatial Dependence of Protein-Water Collective Hydrogen-Bond Dynamics. *Phys Rev Lett* **2013,** *111*, 218101.
30. Heyden, M.; Sun, J.; Funkner, S.; Mathias, G.; Forbert, H.; Havenith, M.; Marx, D. Dissecting the THz Spectrum of Liquid Water from First Principles via Correlations in Time and Space. *P Natl Acad Sci USA* **2010,** *107*, 12068-12073.
31. Wolf, M.; Emmert, S.; Gulich, R.; Lunkenheimer, P.; Loidl, A. Dynamics of Protein Hydration Water. *Phys Rev E* **2015,** *92*, 032727.
32. Lunkenheimer, P.; Emmert, S.; Gulich, R.; Kohler, M.; Wolf, M.; Schwab, M.; Loidl, A. Electromagnetic-Radiation Absorption by Water. *Phys Rev E* **2017,** *96*, 062607.
33. Jungwirth, P. Biological Water or Rather Water in Biology? *J Phys Chem Lett* **2015,** *6*, 2449-2451.
34. Ding, T.; Li, R. Y.; Zeitler, J. A.; Huber, T. L.; Gladden, L. F.; Middelberg, A. P. J.; Falconer, R. J. Terahertz and Far Infrared Spectroscopy of Alanine-Rich Peptides Having Variable Ellipticity. *Opt Express* **2010,** *18*, 27431-27444.
35. Ebbinghaus, S.; Kim, S. J.; Heyden, M.; Yu, X.; Heugen, U.; Gruebele, M.; Leitner, D. M.; Havenith, M. An Extended Dynamical Hydration Shell Around Proteins. *P Natl Acad Sci USA* **2007,** *104*, 20749.
36. Nibali, V. C.; Havenith, M. New Insights into the Role of Water in Biological Function: Studying Solvated Biomolecules Using Terahertz Absorption Spectroscopy in Conjunction with Molecular Dynamics Simulations. *J Am Chem Soc* **2014,** *136*, 12800-12807.
37. Rupley, J. A.; Careri, G. Protein Hydration and Function. *Adv Protein Chem* **1991,** *41*, 37-172.
38. Halle, B. Protein Hydration Dynamics in Solution: a Critical Survey. *Philos T Roy Soc B* **2004,** *359*, 1207-1223.
39. Sirotkin, V. A.; Komissarov, I. A.; Khadiullina, A. V. Hydration of Proteins: Excess Partial Volumes of Water and Proteins. *J Phys Chem B* **2012,** *116*, 4098-4105.
40. Bye, J. W.; Meliga, S.; Ferachou, D.; Cinque, G.; Zeitler, J. A.; Falconer, R. J. Analysis of the Hydration Water around Bovine Serum Albumin Using Terahertz Coherent Synchrotron Radiation. *J Phys Chem A* **2014,** *118*, 83-88.
41. George, D. K.; Charkhesht, A.; Vinh, N. Q. New Terahertz Dielectric Spectroscopy for the Study of Aqueous Solutions. *Rev Sci Instrum* **2015,** *86*, 123105.
42. Vinh, N. Q.; Sherwin, M. S.; Allen, S. J.; George, D. K.; Rahmani, A. J.; Plaxco, K. W. High-Precision Gigahertz-to-Terahertz Spectroscopy of Aqueous Salt Solutions as a Probe of the Femtosecond-to-Picosecond Dynamics of Liquid Water. *J Chem Phys* **2015,** *142*, 164502-7.
43. Cametti, C.; Marchetti, S.; Gambi, C. M. C.; Onori, G. Dielectric Relaxation Spectroscopy of Lysozyme Aqueous Solutions: Analysis of the Delta-Dispersion and the Contribution of the Hydration Water. *J Phys Chem B* **2011,** *115*, 7144-7153.
44. Ellison, W. J. Permittivity of Pure Water, at Standard Atmospheric Pressure, over the Frequency Range 0-25 THz and the Temperature Range 0-100 Degrees C. *J Phys Chem Ref Data* **2007,** *36*, 1-18.
45. Buchner, R.; Hefter, G. Interactions and Dynamics in Electrolyte Solutions by Dielectric Spectroscopy. *Phys Chem Chem Phys* **2009,** *11*, 8984-8999.
46. Oleinikova, A.; Sasisanker, P.; Weingartner, H. What Can Really Be Learned from Dielectric Spectroscopy of Protein Solutions? A Case Study of Ribonuclease A. *J Phys Chem B* **2004,** *108*, 8467-8474.

47. Havriliak, S.; Negami, S. A Complex Plane Representation of Dielectric and Mechanical Relaxation Processes in Some Polymers. *Polymer* **1967,** *8*, 161-210.
48. Fukasawa, T.; Sato, T.; Watanabe, J.; Hama, Y.; Kunz, W.; Buchner, R. Relation Between Dielectric and Low-Frequency Raman Spectra of Hydrogen-Bond Liquids. *Phys Rev Lett* **2005,** *95*, 197802.
49. Buchner, R.; Hefter, G. T.; May, P. M. Dielectric Relaxation of Aqueous NaCl Solutions. *J Phys Chem A* **1999,** *103*, 1-9.
50. Hayashi, Y.; Katsumoto, Y.; Omori, S.; Kishii, N.; Yasuda, A. Liquid Structure of the Urea-Water System Studied by Dielectric Spectroscopy. *J Phys Chem B* **2007,** *111*, 1076-1080.
51. Kaatze, U. Complex Permittivity of Water as a Function of Frequency and Temperature. *J Chem Eng Data* **1989,** *34*, 371-374.
52. Pambou, E.; Crewe, J.; Yaseen, M.; Padia, F. N.; Rogers, S.; Wang, D.; Xu, H.; Lu, J. R. Structural Features of Micelles of Zwitterionic Dodecyl-Phosphocholine (C12PC) Surfactants Studied by Small-Angle Neutron Scattering. *Langmuir* **2015,** *31*, 9781-9789.
53. Markelz, A.; Whitmire, S.; Hillebrecht, J.; Birge, R. THz Time Domain Spectroscopy of Biomolecular Conformational Modes. *Phys Med Biol* **2002,** *47*, 3797-3805.
54. Xu, J.; Plaxco, K. W.; Allen, S. J. Probing the Collective Vibrational Dynamics of a Protein in Liquid Water by Terahertz Absorption Spectroscopy. *Protein Sci* **2006,** *15*, 1175-1181.
55. Yamamoto, N.; Ohta, K.; Tamura, A.; Tominaga, K. Broadband Dielectric Spectroscopy on Lysozyme in the Sub-Gigahertz to Terahertz Frequency Regions: Effects of Hydration and Thermal Excitation. *J Phys Chem B* **2016,** *120*, 4743-4755.
56. Bruggemann, D. A. G. Berechnung Verschiedener Physikalischer Konstanten von Heterogenen Substanzen. *Ann. Phys. Leipzig* **1935,** *24*, 636-664.
57. Garnett, J. C. M. Colours in Metal Glasses and in Metallic Films. *Philos T R Soc Lond* **1904,** *203*, 385-420.
58. Hanai, T. Theory of the Dielectric Dispersion due to the Interfacial Polarization and its Application to Emulsions *Colloid Polym Sci* **1960,** *171*, 23-31.
59. Choy, T. C. *Effective Medium Theory: Principle and Applications*. 1999.
60. Lang, E.; Ludemann, H. D. Pressure and Temperature-Dependence of Longitudinal Proton Relaxation-Times in Supercooled Water to -87 Degrees-C and 2500 Bar. *J Chem Phys* **1977,** *67*, 718-723.
61. Woutersen, S.; Emmerichs, U.; Bakker, H. J. Femtosecond Mid-IR Pump-Probe Spectroscopy of Liquid Water: Evidence for a Two-Component Structure. *Science* **1997,** *278*, 658-660.
62. Fogarty, A. C.; Laage, D. Water Dynamics in Protein Hydration Shells: The Molecular Origins of the Dynamical Perturbation. *J Phys Chem B* **2014,** *118*, 7715-7729.

# Supplemental Information for

# High-Precision Megahertz-to-Terahertz Dielectric Spectroscopy of Protein Collective Motions and Hydration Dynamics

Ali Charkhesht,[1] Chola K. Regmi,[1] Katie R. Mitchell-Koch,[2] Shengfeng Cheng,[1,3] Nguyen Q. Vinh[1,*]

[1] Department of Physics and Center of Soft Matter and Biological Physics, Virginia Tech, Blacksburg, Virginia 24061

[2] Department of Chemistry, Wichita State University, Wichita, Kansas 67260

[3] Macromolecules Innovation Institute, Virginia Tech, Blacksburg, Virginia 24061

* corresponding author: vinh@vt.edu; Phone: 540-231-3158

## Sample preparation

BSA proteins with a molecular weight of 66.1 kDa purchased from Sigma Aldrich (Cat. No. 9048-46-8) were used to prepare BSA solutions. In order to accurately determine the molar concentration as well as the volume filling factor of the protein, $f_p$, BSA proteins were dissolved in 5 ml of deionized water in a volumetric flask. The solutions were prepared by weighing and measuring the volume after dissolving BSA in water for several times to obtain an accurate value of the partial specific volume. The accuracy of these measurements is 0.04 ml or 0.8 %. The temperature of BSA solutions was kept in the range above 0 - 5 °C in an ice box. Before the measurement, solutions were placed outside the ice-box to reach room temperature. We have determined accurately the dielectric response of BSA solutions from 100 MHz to 2.0 THz with concentrations ranging from µM to mM using a MHz-to-THz frequency-domain spectrometer based on a vector network analyzer[1-3] and a THz time-domain spectrometer.[4]

## Dielectric relaxation spectroscopy

The low-frequency dielectric spectra from 100 MHz to 50 GHz of BSA solutions were obtained by using a vector network analyzer (PNA N5225A). The vector network analyzer was combined with a dielectric probe (HP 85070E) and a transmission test set to measure the low-frequency dielectric spectra of solutions. For the dielectric probe measurements, the vector network analyzer was calibrated with air/mercury/water (open/short/load). The cell of the transmission set consists of a coaxial line/circular cylindrical waveguide transition containing solutions. The real and imaginary components of the dielectric response were obtained directly from the system. Solutions were injected into a sample cell made of a large metal body of anodized aluminum and the temperature of solutions was set at 25 °C and controlled with an accuracy of ± 0.02°C using a Lakeshore 336 temperature controller.[2, 5]

**Terahertz frequency-domain dielectric spectroscopy**

We have determined accurately the absorption coefficient, $\alpha$, and the refractive index, $n$, of aqueous solutions using our high-precision terahertz frequency-domain spectroscopy.[1-3] The terahertz spectrometer consists of the above vector network analyzer, frequency multipliers and the matched harmonic detectors from Virginia Diodes, Inc. (Charlottesville, VA), spanning a frequency range from 60 GHz to 1.12 THz. The setup is capable of simultaneously measuring intensity and phase over a large effective dynamical range of fifteen orders of magnitude.[2] Samples were kept in a home-built variable path-length cell with submicron (~80 nm) precision in thickness variation. Temperature of the liquid sample is controlled with the Lakeshore 336 temperature controller. The sample cell was built with a large anodized aluminum to ensure thermal stability of the solutions. The absorption and refractive index of the sample were calculated from the changing of intensity and phase, respectively, as a function of the thickness of aqueous samples. For each frequency, we have taken 200 data points of intensity and phase shift to determine the best fitted values of absorption and refractive index of a solution. The very high dynamic range of the frequency extenders, together with an accurate determination of the sample thickness, allows us to obtain the most precise and accurate MHz-to-THz dielectric spectra reported so far in this frequency range (Figure 1).

Using the above-described spectrometer and sample cell we have measured the change of absorption and phase in a sample as functions of path-length. To mitigate problems associated with multiple reflections of the incident light (standing waves, etalon effect), the thickness of our shortest path-length was selected to be long enough to insure strong attenuation of the incident radiation (transmission $<10^{-2}$). We have determined the absorption coefficients, $\alpha$, and refractive index, $n$, of our samples from linear fits of the change in the absorbance, $\Delta A$, and the unwrapped phase shift, $\Delta\theta_{transmission}$, respectively, with changing path-length, $\Delta l$, as a function of frequency, $v$:

$$\begin{cases} \Delta A = \Delta\left(-\ln[I_{transmission}]\right) = \alpha \cdot \Delta l \\ \Delta(\theta_{transmission}) = \dfrac{n \cdot 2\pi \cdot v}{c} \cdot \Delta l \end{cases} \quad (1)$$

where $I_{transmission}$, $\theta_{transmission}$ are the transmitted intensity and phase, $c$ is the speed of light. This method provides a precise determination of absorption coefficients and refractive indices without the need for precise (and difficult to obtain) measurements of the absolute path-length and the intrinsic optical properties of the sample cell. All experiments were repeated approximately ten times to estimate confidence limits.

### Terahertz time-domain spectroscopy

To obtain the dielectric response at higher frequencies (1.2 - 2 THz), we have employed the THz time-domain spectrometer.[4] The THz time-domain spectrometer provides a broadband THz radiation with real and imaginary parts of the dielectric response using a Ti-sapphire oscillator (82 MHz, 60 fs pulse duration and 1 W). A photoconductive antenna fabricated on a low temperature GaAs substrate (TERA K8, MenloSystems) was used to generate and detect the THz-radiation. The measurements are between 0.5 - 2.0 THz with 10 GHz resolution. Samples were placed at the focal point of the broadband THz radiation. The entire spectrometer is enclosed in a plexiglass box and purged with dry gas prior and during measurements to eliminate water absorption from atmosphere. Solutions were kept in a home-built variable path-length cell with submicron (80 nm) precision in thickness variation. The temperature of solutions was controlled with the previously mentioned temperature controller.

### Measurements of electrical conductivity ($\sigma$)

We have measured the electrical conductivity of aqueous BSA solutions using an impedance analyzer (4294A, Agilent Technology) at 25 °C and 1 kHz. The instrument allows us to measure the electrical conductivity over a large range of BSA concentrations with high accuracy (Figure SI1).

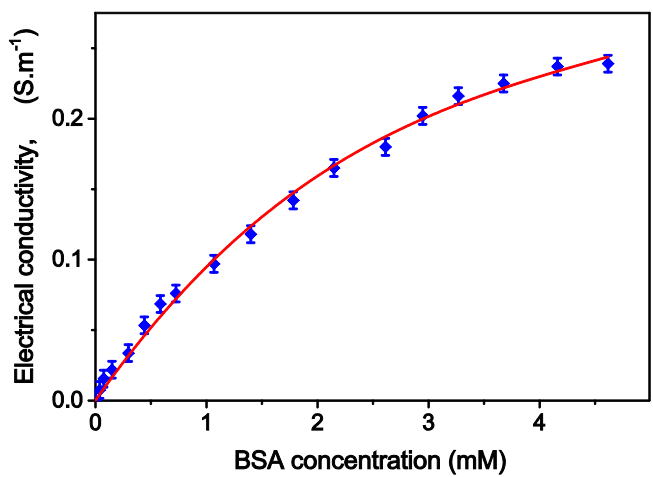

Figure S1: The electrical conductivity, $\sigma$, of aqueous BSA solutions as a function of the protein concentration at 25 °C. The red curve is a guide for eyes.

**References**

1. Vinh, N. Q.; Allen, S. J.; Plaxco, K. W. Dielectric Spectroscopy of Proteins as a Quantitative Experimental Test of Computational Models of Their Low-Frequency Harmonic Motions. *J. Am. Chem. Soc.* **2011,** *133*, 8942-8947.
2. George, D. K.; Charkhesht, A.; Vinh, N. Q. New Terahertz Dielectric Spectroscopy for the Study of Aqueous Solutions. *Rev. Sci. Instrum.* **2015,** *86*, 123105.
3. Vinh, N. Q.; Sherwin, M. S.; Allen, S. J.; George, D. K.; Rahmani, A. J.; Plaxco, K. W. High-Precision Gigahertz-to-Terahertz Spectroscopy of Aqueous Salt Solutions as a Probe of the Femtosecond-to-Picosecond Dynamics of Liquid Water. *J. Chem. Phys.* **2015,** *142*, 164502.
4. Markelz, A.; Whitmire, S.; Hillebrecht, J.; Birge, R. THz Time Domain Spectroscopy of Biomolecular Conformational Modes. *Phys. Med. Biol.* **2002,** *47*, 3797-3805.
5. George, D. K.; Charkhesht, A.; Hull, O. A.; Mishra, A.; Capelluto, D. G. S.; Mitchell-Koch, K. R.; Vinh, N. Q. New Insights into the Dynamics of Zwitterionic Micelles and Their Hydration Waters by Gigahertz-to-Terahertz Dielectric Spectroscopy. *J. Phys. Chem. B* **2016,** *120*, 10757-10767.